\documentclass[letterpaper,prl,twocolumn,showpacs,superscriptaddress]{revtex4}
\usepackage{graphicx}

\begin{document}

\title{Surpassing the standard quantum limit for optical 
imaging using non-classical multimode light}
    
\author{N. Treps} 
\affiliation{Laboratoire Kastler Brossel, Universit\'e Pierre et Marie
Curie, case 74, 75252 Paris cedex 05, France.} 
\affiliation{Department of Physics, Faculty of Science, Australian
National University, Canberra ACT 0200, Australia.} 
\author{U. Andersen} 
\affiliation{Department of Physics, Faculty of Science, Australian
National University, Canberra ACT 0200, Australia.}
\affiliation{Department of Physics, Technical University of Denmark,
DK-2800 Kgs.  Lyngby, Denmark.} 
\author{B. Buchler}
\affiliation{Department of Physics, Faculty of Science, Australian
National University, Canberra ACT 0200, Australia.}
\author{P. K. Lam}
\affiliation{Department of Physics, Faculty of Science,
Australian National University, Canberra ACT 0200, Australia.}
\author{A. Ma\^\i tre}
\affiliation{Laboratoire Kastler Brossel, Universit\'e Pierre et
Marie Curie, case 74, 75252 Paris cedex 05, France.}
\author{H.-A. Bachor}
\affiliation{Department of Physics, Faculty of Science,
Australian National University, Canberra ACT 0200, Australia.}
\author{C. Fabre}
\affiliation{Laboratoire Kastler Brossel, Universit\'e Pierre et
Marie Curie, case 74, 75252 Paris cedex 05, France.}

\begin{abstract}
Using continuous wave superposition of spatial modes, we demonstrate
experimentally displacement measurement of a light beam below the
standard quantum limit.  Multimode squeezed light is obtained by
mixing a vacuum squeezed beam and a coherent beam that are spatially
orthogonal.  Although the resultant beam is not squeezed, it is shown
to have strong internal spatial correlations.  We show that the
position of such a light beam can be measured using a split detector
with an increased precision compared to a classical beam.  This method
can be used to improve the sensitivity of small displacement
measurements.
\end{abstract}

%  Remember to fill in these values
%42.50.-p Quantum optics (for lasers, see 42.55 and 42.60; see also 42.65 Nonlinear optics; 03.65 Quantum mechanics)
%42.50.Dv Nonclassical field states; squeezed, antibunched, and sub-Poissonian states; operational definitions of the phase of the field; phase
%measurements
%42.50.Lc Quantum fluctuations, quantum noise, and quantum jumps
%42.30.-d Imaging and optical processing
%06.30.Bp Spatial dimensions (e.g., position, lengths, volume, angles, displacements)

\maketitle

It has long been known that optical measurements are ultimately
limited in their sensitivity by quantum noise, or shot noise, of the
light.  For more than a decade the usage of nonclassical light has
provided ways of improving the sensitivity beyond this standard
quantum limit \cite{Bachor}.  For example, squeezed light has been
used to improve interferometric \cite{Interf} and absorption
\cite{Absor} measurements.  However, these improvements can only be
applied to signals that correspond to the time modulation of light, as
they rely on the temporal quality of the light.  On the other hand,
many applications require spatial measurement of light.  While
improvements for spatial applications based on non-classical light
have been proposed theoretically \cite{Seng,Fabre}, no experimental
demonstration has yet been shown to work with continuous wave light. 
The challenge is to create strong spatial correlations within a laser
beam, rather than the temporal correlation typically found in
non-classical light sources \cite{kolobov}.  While some experiments
involving sub-Poissonian VCSELS operating in a transverse multimode
regime exhibited a non-random spatial distribution of the quantum
noise \cite{VCSELS}, no spatial correlation was observed within the
produced beam.  Here we present the first successful experimental
demonstration of a spatially ordered light source and a measurement of
the spatial modulation of a laser beam position to below the standard
quantum limit in the continuous wave regime.

This experimental work builds on theoretical work done on
non-classical multimode states of light \cite{Lugiato2}. 
Such states display strong spatial correlations, and their
productions have been the subject of extensive studies in recent years
\cite{kolobov3}.  In particular, the process of parametric
down-conversion in a nonlinear optical medium has been extensively
studied, as it produces ``twin photons'' which are quantum correlated
both temporally and spatially.  Such strong spatial quantum
correlations in the plane perpendicular to the direction of
propagation are produced in spontaneous down-conversion \cite{Teich}
and in multimode transverse optical parametric oscillators
\cite{Lugiato}.  Nevertheless, to our knowledge, there has been no
experimental demonstration of quantum correlations with a multimode
transverse light in the continuous wave regime.

\begin{figure}
    \centerline{\includegraphics[width=8cm]{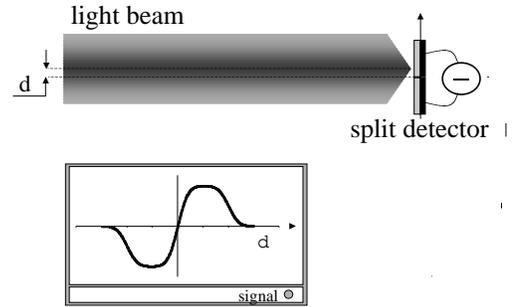}}
\caption{Measurement of the displacement of a light beam.  A split
detector measures the intensities of the two halves of the image
plane.  When the beam is nearly centered, the difference between these
intensities gives a signal that is proportional to the beam
displacement $d$.}
    \label{displacement}
\end{figure}

Precision optical imaging using CCD cameras or photodetector arrays is
required in many areas of science, ranging from astronomy to biology. 
Ultimately, the performance of optical imaging technology is limited
by quantum mechanical effects.  Of particular importance, as far as
applications are concerned, is the measurement of image displacements,
for example, the position of a laser beam.  Techniques that rely on
determining the position of a laser spot include atomic force
microscopy \cite{AFM}, measurement of very small absorption
coefficients via the mirage effect \cite{Boccara} and observation of
the motion of single molecules \cite{biologie}.  These measurements
are usually performed as shown in Fig.~\ref{displacement}.
The beam is incident on a split detector that delivers two currents
proportional to the light intensity integrated over the two halves
($x<0$ and $x>0$) of the image plane.  If the beam is initially
centered on the detector, the mean value of the photocurrent
difference is directly proportional to the relative displacement $d$
of the beam with respect to the detector.  With a classical, shot
noise limited laser source, the smallest displacement that can be
measured (with a signal-to-noise ratio of one) is shown to be
\cite{Fabre}
\begin{equation}
    d_{\rm SQL}=\frac{\sqrt{N}}{2I(0)}.
\end{equation}
Here $N$ is the total number of photons recorded by
the two detectors during the measurement time, and $I(0)$ is the local
density of photons (photons per unit transverse length) at the
position of the boundary between the two detectors.  For a TEM$_{00}$
Gaussian beam with radius $w_{0}$, the minimum measurable displacement is
found to be 
\begin{equation}\label{SQL}
    d_{\rm SQL}=\sqrt{\frac{\pi}{8}}\frac{w_{0}}{\sqrt{N}}.
\end{equation}
For maximum focusing of the Gaussian beam, $w_{0}=\lambda$, and we
obtain $d_{\rm SQL}\approx\lambda/\sqrt{N}$, which is the absolute
minimum displacement of a physical system that can be measured with
classical beams \cite{AFM}.  Equation (\ref{SQL}) shows that a more
powerful laser, or a longer measurement time, gives increased
measurement precision.  However, in many applications these
alternatives are simply not practical.  In the case of atomic force
microscopy, for example, excessive laser power ultimately leads to radiation
pressure noise \cite{FM}.  For biological applications, large laser
power may damage the samples under investigation and an increased 
integration time leads to loss of bandwidth. This is the
motivation for looking for alternative methods of increasing
measurement precision.

\begin{figure}
    \centerline{\includegraphics[width=8.4cm]{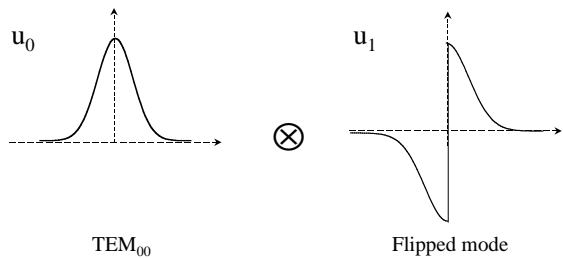}}
    \caption{Electric field profile of the two constituent modes used
    to form the non-classical multimode beam.}
    \label{flipped}
\end{figure}
The limit of equation~(\ref{SQL}) can be surpassed only using
multimode non-classical light.  Let us consider a beam of light with
an electric field distribution given by $E(x)$.  We can build an
orthonormal basis of the transverse plane $\{u_{i}\}$ such that
$u_{0}=E(x)/\parallel E(x)\parallel$ is the first vector; $u_{1}$ is a
``flipped'' mode, given by $-u_{0}(x)$ for $x<0$ and $u_{0}(x)$ for
$x>0$ (see Fig.~\ref{flipped}); and the other modes are choosen in
order to form a basis.  In this basis, the mean field of our light
lies only in the first mode $u_{0}$ but, a priori, all of the modes
contribute to the quantum noise.  In order to determine the relevant
modes of our measurement, we consider the interference quantities
between two modes on each half of the split detector:
\begin{eqnarray}
     I_{x<0}(u_{i},u_{j}) & = & 
     \int_{-\infty}^{0}u_{i}^{*}(x)u_{j}(x)dx \nonumber \\
     I_{x>0}(u_{i},u_{j}) & = & 
     \int_{0}^{+\infty}u_{i}^{*}(x)u_{j}(x)dx.
\end{eqnarray}
Then the interference quantities relevant for a total measurement (sum 
of the two photodetectors) and a differential measurement (difference of 
the two photodetectors) can be written: 
\begin{eqnarray}
    I_{\rm sum}(u_{i}u_{j}) & = & 
    I_{x<0}(u_{i},u_{j})+I_{x>0}(u_{i},u_{j}) \nonumber \\
    I_{\rm diff}(u_{i}u_{j}) & = & 
    I_{x<0}(u_{i},u_{j})-I_{x>0}(u_{i},u_{j})
\end{eqnarray}
One can then show that for any transverse mode $u_{i}$,
 \begin{eqnarray}
%      I_{\rm sum}(u_{i}u_{0}) & = & I_{\rm diff}(u_{i}u_{1}) \nonumber \\
     I_{\rm sum}(u_{i}u_{1}) & = & I_{\rm diff}(u_{i}u_{0}). 
     \label{ortho}
 \end{eqnarray}
Since all $u_{i}$, for $i\geq 2$, are orthonormal to $u_{1}$ (i.e.
$I_{\rm sum}(u_{i}u_{1})=0$), equation~(\ref{ortho}) demonstrates that
these modes
%  
%  This demonstrate that all the transverse modes $u_{i}$, such as 
% $i\geq 2$, as they are orthogonal to $u_{1}$ (i.e. $I_{\rm 
% sum}(u_{i}u_{1})=0$) 
have a zero overlap integral with $u_{0}$ in a differential
measurement.  
% As, for the light beam we are considering, appart from
% $u_{0}$ all the transverse modes are in vacuum states, 
% 
It can then be shown that only $u_{1}$, which has a non-zero overlap
integral with $u_{0}$, has to be considered along with $u_{0}$ in the
noise calculation \cite{Fabre,Delaubert}.
 
We note that the modes $u_{0}$ and $u_{1}$ have perfect interference
visibility as shown by their complete overlap integral for the
differential measurement, ie.  $I_{\rm diff}(u_{0}u_{1})=1$.  In this
regard, the measurement is analogous to a perfect homodyne measurement
with a beam splitter.  The two modes are equivalent to the two input
beams of a beamsplitter and the two halves of the multimode beam are
equivalent to the two outputs.  Therefore, similar to a homodyne
measurement, the noise on the differential measurement is completely
cancelled when the flipped mode is occupied by a perfect squeezed
vacuum, with the squeezed quadrature in phase with the coherent field
of the $u_{0}(x)$ mode.  Conversely, the same result is also obtained
when the mode profiles of the squeezed and the coherent fields are
interchanged.  In order to avoid the effect of losses, we have chosen
a squeezed vacuum mode $u_{0}(x)$.  We would like to stress that this
simplified explanation can be applied only because we have
conveniently identified the two relevant transverse modes of the
measurement.  However, contrary to a homodyne measurement, the entire
measurement is performed using a single beam.  Furthermore, a more
general analysis is not limited to only two-mode beams.

\begin{figure}
    \centerline{\includegraphics[width=8.4cm]{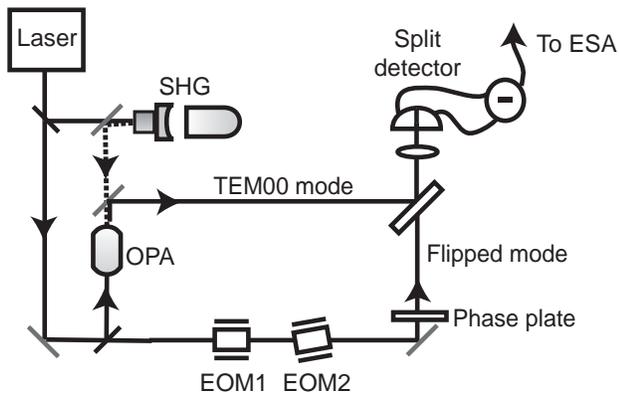}}
    \caption{ Scheme of the experimental setup.  SHG: Second Harmonic
    Generator, OPA: Optical Parametric Amplifier, EOM: Electro-Optic
    Modulator and ESA: Electronic Spectrum Analyzer. The dashed lined 
    correspond to light at 532 nm and the solid line to the light at 
    1064 nm. The TEM00 mode is produced by the OPA and is a squeezed 
    vacuum, the flipped mode is a coherent state.}
    \label{setup}
\end{figure}
The experimental setup is shown in Fig.~\ref{setup}.  A stable Nd:YAG
700~mW laser provides a cw single mode beam at 1064~nm.  A part of
this beam is sent to a locked MgO:LiNbO$_{3}$ frequency doubling
cavity.  The 532~nm output of the frequency doubler is used to pump a
degenerate optical parametric amplifier (OPA) that produces a stable
10~$\mu$W squeezed beam in the TEM$_{00}$ mode at 1064~nm.  The noise
reduction of the OPA output is measured to be 3.5~dB.  Details of this
squeezing system may be found in \cite{Buchler}.  The flipped mode,
$u_{1}(x)$, is produced by sending the remaining part of the initial
1064~nm laser beam through a specially designed phase plate.  This
phase plate consists of two birefringent half-wave plates, one rotated
by 90$^{\circ}$ with respect to the other, forming the two halves
$x<0$ and $x>0$ of the transverse plane.  These elements introduce a
phase shift of 180$^{\circ}$ between the field amplitudes of the two
halves.  The squeezed output from the OPA is required to be
superimposed onto the flipped mode with minimal loss.  This is
achieved by using a beam splitter that reflects 92\% of the squeezed
state and transmits 8\% of the coherent state.  The reflected output
is then sent to a quadrant InGaAs detector (EPITAXX 505Q) with quantum
efficiency greater than 90\%.  Only two of the four quadrants, of
dimensions 500$\mu$m $\times$ 500$\mu$m each and with a dead zone
between the pixels of $25\mu$m, are used in this experiment.  A lens
of focal length 30mm is used to image the phase plate on the detector
plane and to counteract the diffraction of the flipped mode, which
undergoes an abrupt phase change and therefore contains high spatial
frequency components.

\begin{figure}
    \centerline{\includegraphics[width=7.4cm]{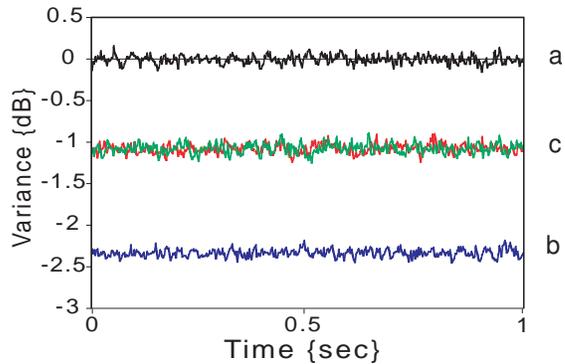}}
    \caption{ Noise spectral density at 4.5 MHz of the photocurrents
    as a function of time (resolution bandwidth 100 kHz).  a) sum of
    the two photocurrents.  b) difference between the photocurrents. 
    c) noise on each detector.}
    \label{spatialsqueezing}
\end{figure}
Fig.~\ref{spatialsqueezing} shows the different noise levels monitored
as a function of time when the relative phase between the coherent
state and the squeezed state is chosen for maximum noise reduction. 
Due to the high stability of the various servo-loops in the
experimental setup, the actively locked operation of the setup can be
kept for hours.  The noise measured on the sum of the two halves
(Fig.~\ref{spatialsqueezing}a), i.e. on the total beam, coincides with
the shot noise level for the conditions of this experiment, as
expected from the coherent beam, which is not affected by the presence
of a squeezed vacuum in an orthogonal mode.  The noise measured on
each individual half (Fig.~\ref{spatialsqueezing}c) is reduced by 1.08
$\pm$ 0.06~dB below the quantum noise limit.  The fact that the
intensity noise on each half of the beam is below the quantum noise
limit, whereas the whole beam is at shot noise, shows the strong
non-classical characteristic of this multimode beam.  This is
corroborated by the experimental data of Fig.~\ref{spatialsqueezing}b,
which gives the noise on the intensity difference between the two
halves at 2.34 $\pm$ 0.05 dB below the quantum limit.  The results
suggest that the beam is made of two strongly quantum correlated
parts, indicating that a significant amount of spatial correlation has
been created among the photons.  With the measured noise reduction in
the squeezed vacuum and a perfect setup (i.e. a perfect phase plate
and a perfect mode-matching between the two transverse modes), one
would expect 2.5 $\pm$ 0.2 dB of noise reduction on the difference
between the two pixels.  This demonstration is, to our knowledge, the
first experiment in which spatial quantum effects have ever been
observed in a bright beam of light.

\begin{figure}
    \centerline{\includegraphics[width=7.4cm]{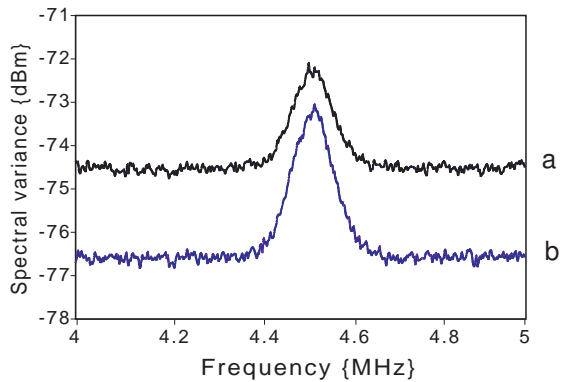}}
    \caption{Noise spectrum of the photocurrent difference in presence
    of an oscillating displacement of amplitude 2.9\AA and frequency
    4.5MHz (resolution bandwidth : 100 kHz).  a) using a coherent
    state of light.  b) using the two-mode non classical state of
    light.  This curve is obtained by averaging the signal over 10
    successive traces.}
    \label{smalldisplacement}
\end{figure}

This spatial noise correlation can now be used to improve the
precision of displacement measurements in the image plane.  For
practical reasons, we have choosen to induce the displacement only in
the coherent mode, before the mixing on the beamsplitter. However
this displacement is of the order of the nanometer, which is several
order of magnitude smaller that the relevant precision for the
mode-matching of the two transverse modes, and the theoretical
prediction for the measurements is the same as if the displacement was
done on the total beam.  In order to produce a small controllable beam
displacement in the frequency range of the previous measurements, we
use two electro-optic modulators (EOMs) driven at 4.5 MHz. 
Fig.~\ref{setup} shows that EOM2 is slightly tilted with respect to
the propagation of the light beam.  When a voltage is applied across
EOM2, a change in refractive index is induced and the transmitted beam
experiences a parallel transverse displacement measured at about
3nm/V. We introduce a modulation at 4.5 MHz as signal for our
displacement measurement which can be easily distinguished from the
low frequency beam displacements induced by mechanical or acoustic
vibrations.  Apart from the parallel displacement, EOM2 will also
introduce an unwanted phase modulation on the transmitted beam which
is detrimental to our measurement.  EOM1 of Fig.~\ref{setup} is
therefore used to compensate for this introduced phase modulation. 
When correct gains are chosen for both modulators, the transmitted
beam will not have any phase or amplitude modulation and is only left
with pure transverse displacement modulation. 
Figure~\ref{smalldisplacement} shows the differential signal monitored
by a spectrum analyzer when the light beam undergoes a displacement
modulation with an amplitude of 2.9\AA. With a resolution bandwidth of
100 kHz, our setup recorded a modulation peak in the Fourier spectrum. 
Fig.~\ref{smalldisplacement}a shows the trace when vacuum instead of
the squeezed vacuum is used in mode $u_{0}(x)$.  Thus this noise floor
gives the standard quantum limit in such a displacement measurement. 
The signal-to-noise ratio (SNR) of this measurement is 0.68.  When the
two-mode non-classical beam is utilized in the measurement
(Fig.~\ref{smalldisplacement}b), we obtain a SNR of 1.20.  This gives
an improvement of the displacement measurement sensitivity by a factor
of 1.7.  The result is in agreement with the theoretical value
calculated with the noise reduction reported in the previous
paragraph.  Similar measurements have been performed with a 10 kHz
resolution bandwidth (and therefore a longer measurement time) and the
results show the same improvement of the SNR.

Our results demonstrate that multimode non-classical states of light can be
utilized to improve the optical measurement of small displacements. 
The noise floor of displacement measurements can actually be reduced
to below the standard quantum limit.  Of particular relevance are the
potential usage of multimode squeezed light in atomic force microscopy
and biological microscopy.  Though our experimental demonstration is
restricted to one-dimensional displacement measurements, it can be
extended to two-dimensional displacement measurements with more
complex forms of multimodal non-classical light.

\begin{acknowledgments}
We would like to thank L. Lugiato and M. Kolobov for many enlightening
discussions and CSIRO, Sydney, for the manufacturing of the special
phase plate.  This work is funded by the European project
IST-2000-26019 "Quantum images", the Centre National de la Recherche
Scientifique and the Australian Research Council.
\end{acknowledgments}


\begin{thebibliography}{12}
\bibitem{Bachor} H-A Bachor, a guide to experiments in quantum optics. 
(Wiley-VCH, Weinheim, 1998).

\bibitem{Interf} M. Xiao, L.A. Wu and H.J. Kimble, Phys.  Rev.  Lett. 
\textbf{59}, 278 (1987); P. Grangier \textit{et al.},
%R.E. Slusher, B. Yurke and A. LaPorta 
Phys.  Rev.  Lett.  \textbf{59}, 19 (1987)

\bibitem{Absor} E. Polzik, J. Carri and H.J. Kimble Appl.  Phys B
\textbf{55}, 279 (1992); F. Marin, A. Bramati, V. Jost and E.
Giacobino, Optics Commun.  \textbf{140}, 146 (1997).

\bibitem{Fabre} C. Fabre, J.B. Fouet and A. Ma\^\i tre, Optics Letters
\textbf{25}, 76-78 (1999).

\bibitem{Seng} Seng-Tiong Ho, P. Kumar and J.H. Shapiro, Phys.  Rev. 
A \textbf{37}, 2017 (1988).

\bibitem{kolobov} M. Kolobov and C. Fabre, Phys.  Rev.  Lett. 
\textbf{85} 3789 (2000).

\bibitem{VCSELS} J.P. Poizat, T. Chang, P. Grangier, Phys.  Rev.  A
\textbf{61}, 043807 (2000); J.P. Hermier \textit{et al.}, 
%A. Bramati, A. Khoury, V. Josse, E. Giacobino, P. Schnitzer, R. Michalzik, K. Ebeling,
IEEE JQE, \textbf{37}, 87 (2001)

\bibitem{kolobov3} M.I. Kolobov, Rev.  Mod.  Phys.  \textbf{71},1539
(1999).

\bibitem{Lugiato2} L.A. Lugiato, A. Gatti and H. Wiedemann in
\textit{Quantum Fluctuations}, Proceedings of the Les Houches Summer
School of Theoretical Physics,
% , Les Houches, 1995, edited by S. Reynaud, E. Giacobino and J. Zinn-Justin 
p431, (North-Holland, Amsterdam, 1997).

\bibitem{AFM} C.A.J. Putman \textit{et al.} 
%B.G. De Grooth, N.F. Van Hulst and J. Greve
, Journ.  of Applied Phys.  \textbf{72}, 6 (1992).

\bibitem{FM} D.P.E. Smith, Rev.  Sci.  Instrum.  \textbf{66}, 3191 (1995).

\bibitem{Boccara} C. Boccara, D. Fournier and J. Badoz, Appl.  Phys. 
Letters \textbf{36}, 130 (1980).

\bibitem{biologie} H. Kojima \textit{et al.},
%E. Muto, H. Higuchi and T. Yanagida, 
Biophysical Journal \textbf{73}, 2012 (1997).

\bibitem{Teich} A. Joobeur \textit{et al.},
%B.E.A. Saleh, T.S. Larchuk and M.C. Teich,
Phys.  Rev.  A \textbf{53}, 5349 (1996).

\bibitem{Lugiato} L.A. Lugiato and I. Marzoli, Phys.  Rev A
\textbf{52}, 4886 (1995); I. Marzoli A. Gatti and L.A. Lugiato, Phys. 
Rev.  Lett.  \textbf{78}, 2092 (1997)

\bibitem{Delaubert} V. Delaubert \textit{et al.},
%D.A. Shaddock, P.K. Lam, B.C. Buchler, H.-A. Bachor and D.E. McClelland 
(submitted to J. Opt.  A, Aug.  2001).

\bibitem{Buchler} B.C. Buchler \textit{et al.}, 
%U.L. Andersen, P.K. Lam, H-A. Bachor, and T.C. Ralph, 
Phys.  Rev.  A, \textbf{65}, 011803(R) (2002).
\end{thebibliography}
\end{document}